# Perspective

# Boosting the discovery of 3D topological materials: mixing chemistry with physics via a two-step computational screening strategy

Xing-Qiu Chen

(Shenyang National Laboratory for Materials Science, Institute of Metal Research, Chinese Academy of Sciences, Shenyang, 110016, P. R. China; Email: xingqiu.chen@imr.ac.cn)

Topological materials in crystal solids, including topological insulators (TIs), topological crystalline insulators (TCIs), topological Dirac semimetals (DSMs), topological Weyl semimetals (WSMs), topological Dirac or Weyl nodal line semimetals (NLSMs) and beyond,[1-5] are mainly featured with topological, protected non-trivial surface states, and their bulk phases are insulators or semimetals with the proper presence of Dirac cones, Weyl nodes or Dirac nodal lines around the Fermi level. Because of their potential for use in spintronics, quantum computers and novel physics, topological materials have attracted interest. Currently, the design of topological materials with appealing properties still heavily relies on a traditional trial-and-error method, despite the achievements that have been made by combining state-of-the-art first-principles calculations and advanced experiments.

Among the three-dimensional (3D) topological materials reported to date, two main features can be observed. The first is that completely filled bands are necessary (Fig. 1a). This feature ensures the insulating behavior of the bulk of TIs or the appearance of semimetals with the possible existence of bulk Dirac cones, Weyl nodes, and nodal lines near the Fermi level in DSMs, WSMs, NLSMs and other semimetals. The second important prerequisite is the occurrence of the so-called band inversion[6] (Fig. 1b), which can be impacted by many factors[1-5], including chemical bonding, crystal field effects, spin-orbit coupling (SOC), strong electron correlations and even van der Waals forces.

Combining these two common features—fully filled bands and band inversion—is a natural approach to developing high-throughput computational screening of topological materials in computational materials science. The first criterion of fully filled bands in materials can be addressed utilizing the simple chemical concept of electronegativity, which is a property that describes the tendency of an atom to attract electrons (or electron density). For instance, the famous 3D TIs ($Bi_2Se_3$ family[7]) are recognized to form fully filled orbitals. As illustrated in Fig. 1c, the composition of $Bi_2Se_3$ satisfies the fully filled $p^6$ orbitals because the Se atom has an electronegativity of 2.55, which is larger than that of Bi (2.02). Using the concept of electronegativity, Se has a $Se^{2-}$ configuration with fully filled $p^6$ orbitals by obtaining extra electrons from Bi, which has a $Bi^{3+}$ configuration. In contrast, Bi may exhibit a $Bi^{3-}$ configuration by obtaining three electrons to fully fill the $6p^6$ orbital. This requires that the electronegativity of the other counterparts in the materials is smaller than that of Bi. As illustrated in Fig. 1c, because the electronegativity of the alkali metals is smaller than that of Bi, it can form a fully filled $6p^6$ orbital in $Na_3Bi$ by transferring the Na $3s^1$ electron to Bi. This is true, and $Na_3Bi$ was the first experimental realization of DSMs[8,9] after our prediction[4] in collaboration with Professor Zhong Fang's group at the Institute of Physics, Chinese Academy of Science, which led to important progress, both theoretically and experimentally. Similar to $Na_3Bi$, the $Sr_3Bi_2$ family was theoretically demonstrated to be wide-gap TIs[10] because the alkali-earth metals have a smaller electronegativity value than Bi. Of course, it needs to be emphasized that the so-called fully filled bands not only denote the bands from atomic orbitals with full shell structure (such as 8-electron or 18-electron rules), but also involve the bands from sub-group of atomic orbitals well separated by crystal field, such as fully filled $d$-$t_{2g}$ orbitals in cubic crystal field.

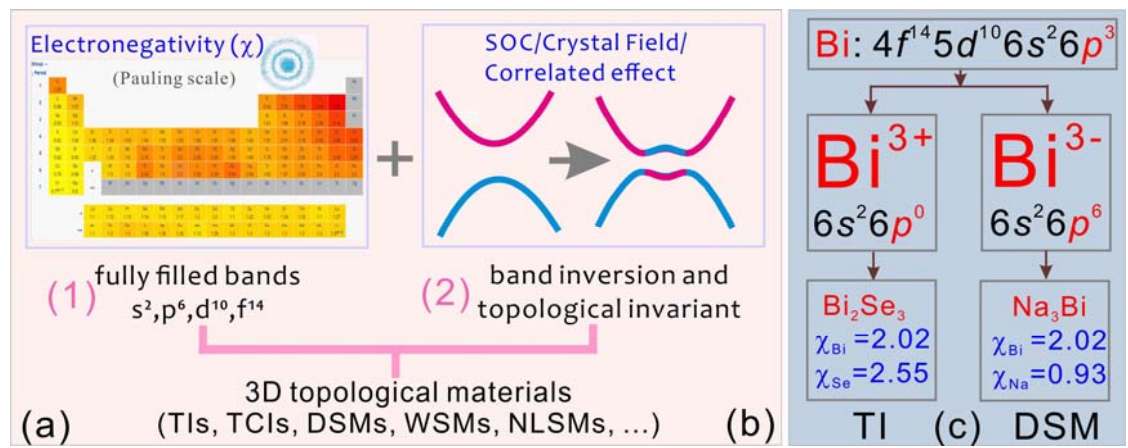

**Figure 1 Schematic of a two-step computational screening strategy.** The left panels denote the concepts of fully filling the bands (a) according to the electronegativity in the periodic table of elements and of analyzing the band inversion (b) in association with the topological states. The right panels (c) show the cases of a famous TI ($Bi_2Se_3$ family) and DSM ($Na_3Bi$ family) from the first computational screen with electronegativity.

Certainly, the first electronegativity screening strategy for the fully filled bands is not a final answer as to whether the screened materials are topological. The main purpose is to narrow the wide search down to a manageable window for potential candidates. This narrowing will heavily reduce the time-consuming aspect of computational design. By using key empirical knowledge—*i.e.*, some heavy metallic elements, such as Bi, Pb, Tl and Hg, have a strong SOC and possibly induce the band inversion by interacting with the crystal field effects—in combination with the electronegativity considerations, the search can be more effective. It should be emphasized that the strong SOC is not the only important factor that causes the topological states. For instance, the topological states in TCIs (*e.g.*, rock-salt SnTe[11] and SnSe[12,13]) are protected by the mirror symmetry of the lattice (rather than SOC), and one type of Dirac NLSMs requires the SOC strength to be as small as possible, such as pure alkali-earth, fully filled *s*-orbital beryllium.[14] Therefore, once the first screen is performed, the second screening strategy is extremely important to physically identify the topological states of a targeted material and to analyze the topologically protected surface states. The second screen needs to calculate the topological invariants ($Z$ or $Z_2$)[15]. For a lattice with an inversion symmetry, the $Z_2$ invariant can be derived by calculating the parities of all occupied bands at eight time-reversal invariant momenta (TRIMs) in the Brillouin zone (BZ)[16]. However, no matter whether lattice has inversion center the Wilson loop method is useful to identify topological property by elucidating Wannier center evolution loop of each occupied band in the TRIMs-contained planes of the BZ.[17,18] The $Z_2$ invariant is related to the summation of winding numbers on the cylinder surface along the loop between two TRIM points for all paired Wannier centers. The non-trivial topological property can be identified by a count of how many times (even or odd) the evolution lines cross an arbitrary reference line[18].

The ongoing efforts to discover topological materials involve developing a high-throughput computational code to design topological materials via combining two-step computational screens within the frameworks of the Big Data, Materials Informatics, and Materials Genome Initiative. The core challenge is to elucidate a high-throughput, fast, accurate algorithm to boost the predictive power for the topological states among the various realistic complexities of topological materials.

**Acknowledgements:** Work was supported by the Hundred Talents Project of the Chinese Academy of Sciences, the National Natural Science Foundation of China (Grant No. 51671193) and the Science Challenging Project No. TZ2016004.